\newcommand\lsim{\mathrel{\rlap{\lower4pt\hbox{\hskip1pt$\sim$}}
    \raise1pt\hbox{$<$}}}
\newcommand\gsim{\mathrel{\rlap{\lower4pt\hbox{\hskip1pt$\sim$}}
    \raise1pt\hbox{$>$}}}
\documentstyle[prl,floats,epsfig,aps,twocolumn]{revtex}

\begin{document}
\twocolumn[\hsize\textwidth\columnwidth\hsize\csname @twocolumnfalse\endcsname
\title{The growth of linear perturbations in generic defect models for structure formation}

\author{P.\ P.\ Avelino$^{1}$ and J.\ P.\  M.\ de~Carvalho$^{1,2}$}
\address{$^{1}$Centro de Astrof{\' \i}sica, Universidade do Porto, Rua das
Estrelas s/n,
4150 Porto, Portugal}
\address{$^{2}$Dep.\ de Matem\'atica Aplicada da Faculdade de Ci\^encias da Univ.\ do Porto,
Rua das Taipas 135, 4050  Porto -- Portugal}

\maketitle

\begin{abstract}

We study the growth of linear perturbations induced by a generic causal scaling source as a 
function of the cosmological parameters $h$, $\Omega^{\rm m}_0$ and $\Omega^\Lambda_0$. We show that 
for wavenumbers $k \gsim 0.01 \, h \, {\rm Mpc}^{-1}$ the spectrum of density and velocity perturbations 
scale in a similar way to that found in inflationary models with primordial perturbations. 
We show that this result is independent of the more or 
less incoherent nature of the source, 
the small scale power spectrum of the source and of deviations 
from scaling which 
naturally occur at late times if $\Omega^{\rm m}_0 \neq 1$. 

\end{abstract}
\vskip .3in 
]


 
\subsection{\bf Introduction} 
\label{intro} 
 
At present there are two main competing paradigms for the origin of the large scale structure  
of the Universe: topological defects and inflation. In defect scenarios\cite{VilShe}  
density perturbations are  
causally seeded by defect evolution on increasingly larger scales while in inflationary models  
the causal structure of the universe is altered and correlations can be established on scales  
much larger than the Hubble volume. Inflationary perturbations are produced at a remote epoch  
and their evolution is linear until late times. They are passive and coherent.  
In contrast the evolution of topological defects is highly non-linear  
and can only be fully described resorting to large numerical simulations\cite{AllShe,BenBou2}.  
Defect perturbations are  
active and more or less incoherent depending on the source. Despite the differences between them  
both scenarios produce a nearly scale invariant spectrum of density perturbations.  
Hybrid models featuring both inflationary and defect perturbations have also been  
proposed\cite{AveCal3,ConHin1,BatWel}. 
 
 
Recently there has been substantial progress in the understanding of defect models as seeds for the large scale structure and cosmic microwave background anisotropies 
\cite{AveShe2,ACDKSS,AveShe4,against,ConHin,PenSel}. 
The matter and microwave anisotropy power spectrum induced by local and global strings, 
global monopoles and global textures have been calculated 
and compared with current observations. Although the results seem to indicate some 
disagreement between standard defect models and observations (in particular in what concerns the cosmic microwave background anisotropies), some issues still remain to be 
investigated in more detail. This is particularly true 
for local strings in which the exact contribution of cosmic string loops 
and gravitational radiation is still uncertain\cite{AveCal2,AveShe5}. 
 
Given that current observational data strongly suggest that the density of the Universe is sub-critical  
(favoring a non-zero cosmological constant)\cite{Per,Lin} it is very important that structure formation analysis  
performed for defect models can be generalized for any realization of the cosmological  
parameters $h$, $\Omega^{\rm m}_0$ and $\Omega^\Lambda_0$. This has been done in particular for  
cosmic strings\cite{AveShe2,AveCal1,BRA}  but the validity of the approximations used have not been investigated 
for generic  
defect structure formation scenarios. In this letter we describe the dependence of  
the power spectrum of density and velocity perturbations, induced by a 
generic  
causal scaling source, on the cosmological parameters $h$, 
$\Omega^{\rm m}_0$ and $\Omega^\Lambda_0$. 
We show that for $k \gsim 0.01 \, h \, {\rm Mpc}^{-1}$ these scale in a very similar way to  
that found in inflationary models with primordial fluctuations. We also study the deviations  
from scaling which naturally occur at late times if $\Omega^{\rm m}$ starts deviating from unity.  
This is modeled by switching off the source when $\Omega^{\rm m}$ reaches a certain fixed critical  
value $\Omega^{\rm m}_* < 1$. We demonstrate that our results hold independently of the more or less incoherent  
nature of the source, the small scale power spectrum of the source and of deviations  
from scaling which naturally occur at late times if $\Omega^{\rm m}_0 \neq 1$\cite{Mar}. Finally, we generalize our results for an arbitrary red-shift.
 
\subsection{The growth of linear perturbations}  
\label{growth}

The dynamical equation which describes the evolution of the scale factor $a$ in a 
Friedmann-Robertson-Walker (FRW) universe  
containing both cold dark matter (CDM) and 
radiation fluids can be written as: 
\begin{equation}   
{\cal H}^2=(\Omega^{\rm m}_0 a^{-1} + \Omega^{\rm r}_0 a^{-2}+\Omega^\Lambda_0 a^2 + \Omega^{\rm k}_0)\eta_{\rm eq}^{2}. 
\label{one} 
\end{equation} 
where ${\cal H}={\dot a / a}$ (a dot represents a derivative with respect 
to $\tau=\eta/\eta_{\rm eq}$ and $\eta$ is the conformal time), 
$\Omega^{\rm m}$, 
$\Omega^{\rm r}$ and $\Omega^\Lambda$ express the densities in CDM, radiation and cosmological  
constant as fractions of the 
critical density and $\Omega^{\rm k} = {1 - \Omega^{\rm m}- \Omega^{\rm r} -\Omega^\Lambda}$. The subscript 
`$0$' means that the quantities are to be evaluated today and the subscript `${\rm eq}$' 
means that the quantities are to be evaluated at the time of equal matter and radiation. We have also assumed that 
$a_0=1$ and ${\cal H}(\tau_0)=\eta_{\rm eq}$. 
  
The evolution of radiation and CDM density 
fluctuations ($\delta^{\rm r} =\delta \rho^{\rm r}/\rho^{\rm r}$ and 
$\delta^{\rm m}=\delta \rho^{\rm m}/\rho^{\rm m}$ 
respectively) in the synchronous gauge is given by\cite{VS90}: 
\begin{eqnarray}  
{{\ddot {\cal S}}^{\rm m}}_{\bf k}  + {\cal H} {\dot {\cal S}}^{\rm m}_{\bf k} - {3 \over 2}{\cal H}^2 
(\Omega^{\rm m} {\cal S}^{\rm 
m}_{\bf k} +  2 \Omega^{\rm r} {\cal S}^{\rm m}_{\bf k}) = {\cal D}({\bf k},\tau), 
\label{two} 
\end{eqnarray} 
\begin{equation} {{\ddot {\cal S}}^{\rm r}}_{\bf k}  + {1 \over 3} k^2 {{\cal S}}^{\rm r}_{\bf k}   - {4 \over 3} 
{{\ddot {\cal S}}^{\rm m}}_{\bf k}  = 0, 
\label{three} 
\end{equation} 
where ${{\cal S}}^{\rm r,m}_{\bf k}=\delta^{\rm r,m}_{\bf k} k^{3 \over 2}$, 
$\delta^{\rm r,m}_{\bf k}$ is  
the Fourier transform of 
$\delta^{\rm r,m}$, 
$k=|{\bf k}|$ is given in units of 
$(c\eta_{\rm eq})^{-1}$, ${\cal D}({\bf k}, \tau) = 4 
\pi k^{3/2} G \Theta_+$, $\Theta_+=\Theta_{00}+\Theta_{ii}$ and 
$\Theta_{\alpha\beta}$ is the energy-momentum tensor of the external 
source. 
 
At early times when $a \ll {\rm min}((\Omega^{\rm m}_0/ \Omega^\Lambda_0)^{1 \over 3},\Omega^{\rm m}_0/ \Omega_0^{\rm k})$ the scale factor 
is a quadratic function of the conformal time: 
\begin{equation}   
a(\tau) = a_{\rm eq} \left( 2(\sqrt{2} - 1) {\tau} + (3 - 2 \sqrt{2}){\tau^2} \right), \label{four} 
\end{equation} 
and  
\begin{equation}  
\Omega^{\rm m}={a \over {a+a_{\rm eq}}}, \qquad \Omega^{\rm r}={a_{\rm eq} \over {a+a_{\rm eq}}}, 
\label{five} 
\end{equation} 
with $a_{\rm eq} \propto (\Omega^{\rm m}_0 h^2)^{-1}$. 
For a scaling source no other length scale apart from the horizon scale is involved and so 
\begin{equation}  
{\cal D}({\bf k}, \tau) = \phi^2_0 k^{3/2}\tau^{-1/2}{\cal F}^{1 \over 2}({\bf k}, \tau), 
\label{six} 
\end{equation} 
with 
\begin{equation}  
\langle |{\cal D}({\bf k}, \tau)|^2 \rangle = \phi^4_0 k^3\tau^{-1}
{\cal F}_{\rm av}(k \tau), 
\label{seven} 
\end{equation} 
due to homogeneity and isotropy. Here $\phi_0$ is the defect symmetry breaking scale.
 
We can see from equations (\ref{one}-\ref{six}) that for a scaling 
source the evolution of density 
perturbations on a wavenumber $k$ up to a conformal time $\tau_f$ 
satisfying $a(\tau_f) \ll {\rm min}((\Omega^{\rm m}_0/ \Omega^\Lambda_0)^{1 \over 3},\Omega^{\rm m}_0/  
\Omega_0^{\rm k})$ does not depend on the cosmological parameters $h$, 
$\Omega^{\rm m}_0$ and $\Omega^\Lambda_0$ (note that $k$ is given in units of 
$(c\eta_{\rm eq})^{-1} \propto \Omega^{\rm m}_0 h^2 \, {\rm Mpc}^{-1}$ 
and $\tau=\eta/\eta_{\rm eq}$).  
We note that this conclusion still holds in the more realistic case in which the right hand 
side of equation (\ref{six}) is multiplied  
by a function of $\tau$ to account for the deviation from scaling which naturally occurs 
during the radiation-matter transition. 
Consequently, we conclude that in order 
to study the dependence of the normalization of the power spectrum of 
density fluctuations, 
induced by a scaling source, on the  
cosmological parameters 
$h$, $\Omega^{\rm m}_0$ and $\Omega^\Lambda_0$ we only need to study the  evolution of 
perturbations from a conformal time $\tau_i$ onwards  
(with $\Omega^{\rm r}_0/ \Omega^{\rm m}_0 \ll 
a(\tau_i) \ll {\rm min}((\Omega^{\rm m}_0/ \Omega^\Lambda_0)^{1 \over 3},\Omega^{\rm m}_0/  
\Omega_0^{\rm k})$). In this case equation (\ref{two}) reduces to: 
\begin{equation}  
{{\ddot {\cal S}}^{\rm m}}_{\bf k}  + {\cal H} {\dot {\cal S}}^{\rm m}_{\bf k} - 
{3 \over 2} \eta_{\rm eq}^2 \Omega^0_{\rm m} {\cal 
S}^{\rm m}_{\bf k}/a = {\cal D}({\bf k},\tau). 
\label{eight} 
\end{equation} 
We solve this equation numerically 
assuming a simple form for the  scaling source: 
\begin{equation}  
{\cal F} ({\bf k}, \tau)={\cal F} (k \tau) =C \phi_0^4 \Theta(\beta k \tau-1)\Theta(1-\beta_{*} k \tau), 
\label{nine} 
\end{equation} 
with $\beta / \beta_* \ge 1$, $C = constant$ and $\Theta(x)$ is the step function ($\Theta(x)=0$ if $x \le 0$ and $\Theta(x)=1$ if $x \ge 0$). The initial conditions in equation 
(\ref{eight}) are ${\cal S}^{\rm m}_{\bf k}(\tau_i)=0$ and 
${\dot {\cal S}}^{\rm m}_{\bf k}(\tau_i)=0$ for 
$\beta k \tau_i \le 1$.
This source incorporates the most important features of  
causal scaling models relevant to our study. 
The parameter $\beta$ is related to the compensation scale while $\beta_{*}$ is a small scale cut-off which models the deviation of the power 
spectrum of the source from a white-noise spectrum on small scales. 
For the moment we are 
assuming that the source is coherent but in the next section we will see that this does not  
affect our results. 
 
Different values of $\Omega^{\rm m}_0$ and $\Omega^\Lambda_0$ lead to different linear growth  
factors from early times to the present. For primordial perturbations 
the quantity $\Omega^{\rm m}_0 h^2  
g(\Omega^{\rm m}_0,\Omega^\Lambda_0)$ where 
\begin{equation}  
g(\Omega^{\rm m}_0,\Omega^\Lambda_0)={{{5 \Omega^{\rm m}_0}/2} \over {(\Omega^{\rm m}_0)^{4/7}  
-\Omega^\Lambda_0+(1+\Omega^{\rm m}_0/2)(1+\Omega^\Lambda_0/70)}}
\label{ten}  
\end{equation}  
provides a very good fit to the growth factor of density perturbations from early times 
($\Omega^{\rm r}_0/ \Omega^{\rm m}_0 \ll a(\tau_i) \ll 
{\rm min}((\Omega^{\rm m}_0/ \Omega^\Lambda_0)^{1 \over 3},\Omega^{\rm m}_0/  
\Omega_0^{\rm k})$) to the present\cite{CarPre} (see also\cite{Eis}). This fit was shown to be 
good to a  
few percent for $0.1 \le \Omega^{\rm m}_0 \le 1$ and $0 \le \Omega^\Lambda_0 \le 1$ and 
is normalized to unity for $\Omega^{\rm m}_0=1$ and  
$\Omega^\Lambda_0=0$.

 
A quantity directly related to the amplitude of matter perturbations is their rate of growth which may be described by :
\begin{equation}  
f \equiv {{d \ln \delta} \over {d \ln a}}.
\label{eleven} 
\end{equation}  
For primordial perturbations a very accurate approximation to the present rate of growth 
of matter 
perturbations 
is given by\cite{LahLil}
\begin{equation}  
f(\Omega^{\rm m}_0,\Omega^\Lambda_0)=({\Omega^{\rm m}_0})^{0.6} + {\Omega^\Lambda_0 \over {70}}\left(1+{\Omega^{\rm m}_0 \over 2}\right). 
\label{twelve} 
\end{equation}  
In the next section we will show how these results can be generalized for generic defect models of structure formation. 
 
\begin{figure}
\centerline{\epsfig{file=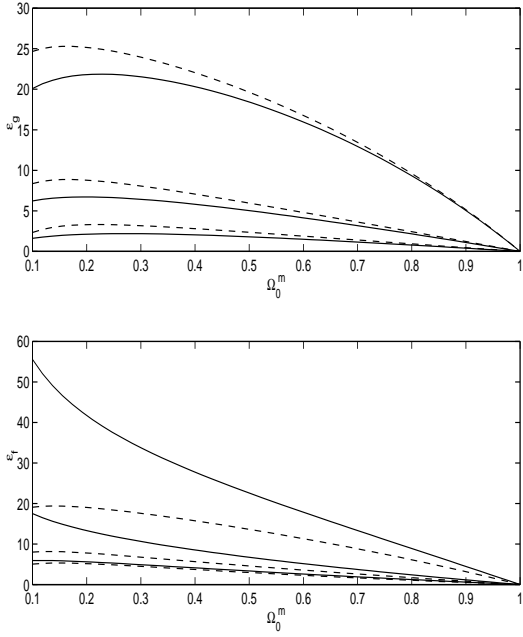,width=7.0cm}}
\caption{Plot of $\epsilon_g$ and $\epsilon_f$ as a function of 
$\Omega_0^{\rm m}$ for open universes 
(solid lines) and flat universes (dashed-lines) in the case with 
$\beta / \beta_* \to \infty$ for $k=0.05,0.01$ and $0.002 \, h \, {\rm Mpc}^{-1}$ in ascending order.}
\label{figefold}
\end{figure}

\subsection{\bf Results and discussion} 
\label{biaf} 
 
The power spectrum induced by a scaling source for arbitrary values 
of the cosmological parameters $h$, $\Omega_0^{\rm m}$ and 
$\Omega_0^\Lambda$ can 
be written as
\begin{equation}  
S_k(h,\Omega^{\rm m}_0,\Omega_\Lambda^0)=A \cdot  
g_k^2(\Omega^{\rm m}_0,\Omega^\Lambda_0) \cdot  
(\Omega_0^{\rm m}h^2)^2 \cdot S_k(1,1,0),
\label{thirteen} 
\end{equation} 
where 
\begin{equation}
S_k(h,\Omega^{\rm m}_0,\Omega^\Lambda_0)=\langle 
|{\cal S}_{\bf k}(h,\Omega^{\rm m}_0,\Omega^\Lambda_0,\tau_0)|^2 
\rangle,
\label{thirt1} 
\end{equation} 
$A={{\phi_0^4(\Omega^{\rm m}_0,\Omega^\Lambda_0)} / {\phi_0^4(1,0)}}$ and 
$k$ is given in units of $(c \eta_{\rm eq})^{-1}$ (or equivalently, in units 
of $\Omega^{\rm m}_0 h^2 \, {\rm Mpc}^{-1}$). The factor $A(\Omega^{\rm m}_0,\Omega^\Lambda_0)$ can be normalized using, for example, the cosmic microwave 
background observations.

It is also useful to define 
\begin{equation}  
f_k \equiv {{d \ln  |{\cal S}_{\bf k}|} \over {d \ln a}},
\label{fourteen} 
\end{equation}  
in a similar way to equation (\ref{eleven}).

The functions 
\begin{equation}  
\epsilon_g(k) \equiv |1-g_k/g| \times 100,
\label{fourt1} 
\end{equation} 
and
\begin{equation}
\epsilon_f(k) \equiv |1-f_k/f| \times 100,
\label{fourt2} 
\end{equation} 
measure the deviation of 
$f_k$ and $g_k$ from $f$ and $g$, given respectively by 
equations (\ref{ten}) and (\ref{twelve}), evaluated at the present time.


In figure \ref{figefold} we plot $\epsilon_g$ and 
$\epsilon_f$ as a function of 
$\Omega_0^{\rm m}$ for open universes 
(solid lines) and flat universes (dashed-lines) in the case with 
$\beta / \beta_* \to \infty$. 
We can see that $g_k = g$ and $f_k = f$ for 
$k \gsim 0.01 \, h \, {\rm Mpc^{-1}}$ to high accuracy. We note that the case 
with $\beta / \beta_* \to \infty$ is the worst possible scenario as far as 
the smallness of $\epsilon_g$ and $\epsilon_f$ is concerned. For 
$\beta / \beta_* \lsim 2$, for example,  $g_k \sim g$ and $f_k \sim f$ 
to better than $5 \%$ for $k \gsim 0.003 \, h \, {\rm Mpc^{-1}}$.

Here we have chosen a specific value for 
$\beta$ ($\beta=1/{\sqrt {18}}$) valid for the matter era and 
motivated by the results of ref.\cite{CheMag}. The change 
in $\beta$ which occurs at late times if $\Omega^0_{\rm m} \neq 1$, 
does not alter our results.
We have also verified that our results are 
not modified if we replace our sharp cut-off on large scales in equation 
(\ref{nine}) by the more realistic cut-off $1/(1+(k\beta \tau)^2)^2$.  

\begin{figure}
\centerline{\epsfig{file=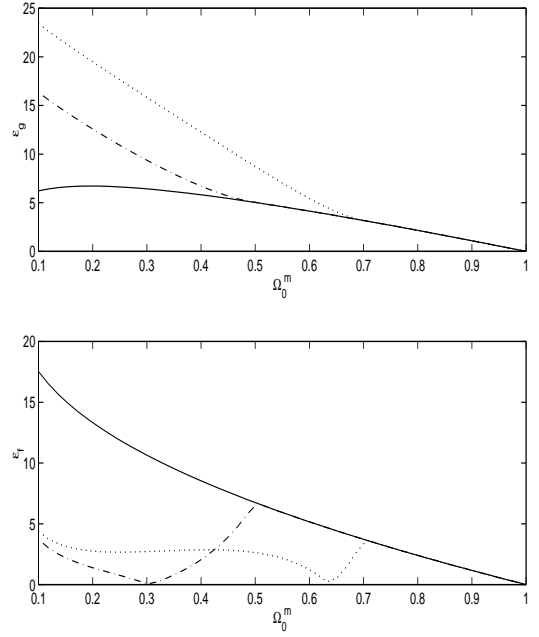,width=7.0cm}}
\caption{Plot of $\epsilon_g$ and $\epsilon_f$ as a function of 
$\Omega_0^{\rm m}$ for an open universe in the case with 
$\beta / \beta_* \to \infty$ for $k=0.01 \, h \, {\rm Mpc}^{-1}$ and 
$\Omega^{\rm m}_*=0,0.5,0.7$ (solid, dot-dashed and dotted lines, 
respectively).}
\label{figefold1}
\end{figure}

We also investigate the effect of deviations from scaling which 
may occur at late times by switching off the source for 
$\Omega^{\rm m} \leq \Omega^{\rm m}_* < 1$. In figure \ref{figefold1} we 
plot $\epsilon_g$ and $\epsilon_f$ as a function of 
$\Omega_0^{\rm m}$ for open universes in the case with 
$\beta / \beta_* \to \infty$ for $k=0.01 \, h \, {\rm Mpc}^{-1}$ and 
$\Omega^{\rm m}_*=0,0.5,0.7$. We can see that our results are not greatly 
influenced by the late time behavior of the source. 
This is specially true if $\beta / \beta_* \lsim 5$ or for flat universes, 
in which case  $\epsilon_{g,f} \lsim 0.05$ for $\Omega^{\rm m}_* \lsim 0.9$ 
and $k \gsim 0.01 \, h \, {\rm Mpc}^{-1}$.

Using the linearized continuity equation 
\begin{equation}  
{\dot \delta}_{\bf k}=-i{\bf k} \cdot {\bf v},
\label{fifteen} 
\end{equation}
where ${\bf v}$ is the peculiar velocity, it can also be inferred 
from our results that 
the velocity spectrum at the present time 
can also be rescaled for arbitrary values of the 
cosmological parameters $h$, $\Omega_0^{\rm m}$ and $\Omega_0^\Lambda$ in 
a simple and accurate way:
\begin{equation} 
{\bf v}_{\bf k}(\Omega^{\rm m}_0, \Omega^\Lambda_0)=h \cdot f(\Omega^{\rm m}_0, \Omega^\Lambda_0) \cdot 
g(\Omega^{\rm m}_0, \Omega^\Lambda_0) \cdot {\bf v}_{\bf k}(1,0),
\label{sixteen} 
\end{equation}
for $k \gsim 0.01 \, h \, {\rm Mpc}^{-1}$. 

We have also verified that our results hold for an incoherent source 
satisfying :
\begin{equation}  
\langle \Theta_+({\bf k}, \tau) \Theta_+(-{\bf k}, \tau+\Delta \tau) 
\rangle = 0,
\end{equation} 
for $\Delta \tau > \tau_{\rm c}$ where $\tau_{\rm c}$ is assumed to be 
smaller than one Hubble time and
\begin{equation}  
{\cal F}(k \tau) \propto 
\int^{\tau_{\rm c}}_{-\tau_{\rm c}} \langle \Theta_+({\bf k}, \tau) 
 \Theta_+({\bf -k}, \tau+\tau') \rangle d \tau',
\end{equation} 
is given by equation (\ref{nine}). 

Finnaly we note that our results may be generalized for an arbitrary red-shift $z = 1/a-1$:
\begin{equation}  
S_k(h,\Omega^{\rm m},\Omega^\Lambda,z)= {S_k(h,\Omega^{\rm m}_0,\Omega^\Lambda_0) \over 
{(1+z)}^2} \cdot  
{{g^2(\Omega^{\rm m},\Omega^\Lambda)} \over {g^2(\Omega^{\rm m}_0,\Omega^\Lambda_0)}}, 
\end{equation}
for $k \tau \gsim 0.02$.

Although we have assumed the dark matter to be cold our results are also valid for hot 
dark matter due to scaling of the neutrino free-streaming length with 
$\Omega^{\rm m}_0 h^2$ (see for example\cite{AveCar}).

 
\subsection{\bf Conclusion} 
\label{conc}

In this letter we have shown that the spectrum of density and velocity perturbations  
induced by a causal scaling source for a particular realization of the cosmological  
parameters $h$, $\Omega^{\rm m}_0$ and $\Omega^\Lambda_0$ can be generalized in a very simple and 
accurate way for any reasonable choice of these parameters on scales $k \gsim 
0.01 \, h \, {\rm Mpc}^{-1}$. 
This rescaling greatly simplifies the study of structure formation with 
topological defects because it allows for an easy generalization, 
on all scales of cosmological interest,  of the 
results of structure formation simulations induced by causal scaling sources 
for any reasonable values of the comological parameters $h$, 
$\Omega^{\rm m}_0$ and $\Omega^\Lambda_0$.
We have verified that our results  
are generic in the sense that they hold independently of 
the more or less incoherent  
nature of the source, the small scale power spectrum of the source and of 
deviations from scaling which may occur at late times.

 
\acknowledgements 
 
P.P.A. is funded by JNICT (Portugal) under the `Program PRAXIS XXI' (grant no. PRAXIS 
XXI/BPD/9901/96). We would like to  
thank Joana Almeida, Robert Caldwell, Carlos Martins, Paulo Macedo, 
Ant\'onio Morais,  
Paul Shellard, Pedro Viana, Jiun-Huei 
Wu for useful discussions. We thank Centro de Astrof{\' \i}sica da Universidade do Porto (CAUP) for the 
facilities  provided. 
 


 

\begin{references} 
\vspace{-.5in} 
 
\bibitem{VilShe} 
For a review see A.\ Vilenkin and E.\ P.\ S.\ Shellard,  
{\it Cosmic strings and other topological defects}  
(Cambridge University Press, 1994). 
 
\bibitem{AllShe}  
B.\ Allen and E.\ P.\ S.\ Shellard, 1990, 
{\bf Phys.\ Rev.\ Lett.}, {\bf 64}, 685. 
 
\bibitem{BenBou2} 
D.\ P.\ Bennett and F.\ R.\ Bouchet, 
{\rm Phys.\ Rev.\ D} {\bf 41}, 2408 (1990). 
 
\bibitem{AveCal3} 
P.\ P.\ Avelino, R.\ R.\ Caldwell, and C.\ J.\ A.\ P.\ Martins  
{\rm Astro-ph/9809130}.

\bibitem{ConHin1} 
C.\ Contaldi, M.\ Hindmarsh and J.\ Magueijo, {\rm 
Astro-ph/9809053}. 

\bibitem{BatWel} 
R.\ A.\ Battye and J.\ Weller, {\rm Astro-ph/9810203}.
 
\bibitem{AveShe2} 
P.\ P.\ Avelino, E.\ P.\ S.\ Shellard, J.\ H.\ P.\ Wu, B.\ Allen,  
{\rm Phys.\ Rev.\ Lett.} {\bf 81}, 2008 (1998).
 
\bibitem{ACDKSS} 
B.\ Allen, R.\ R.\ Caldwell, S.\ Dodelson, L.\ Knox, E.\ P.\ S.\ Shellard,  
and A.\ Stebbins, {\rm Phys.\ Rev.\ Lett.} {\bf 79}, 2624  (1997). 
 
\bibitem{AveShe4} 
P.\ P.\ Avelino, E.\ P.\ S.\ Shellard, J.\ H.\ P.\ Wu and B.\ Allen, {\rm 
Astro-ph/9803120} ({\rm Ap.\ J.\ Lett.} in press).
 
\bibitem{against} 
A.\ Albrecht, R.\ A.\ Battye, and J.\ Robinson, {\rm Phys.\ Rev.\ Lett.}  
{\bf 79}, 4736 (1997). 
 
\bibitem{ConHin} 
C.\ Contaldi, M.\ Hindmarsh \& J.\ Magueijo, {\rm 
Astro-ph/9808201}. 
 
\bibitem{PenSel} 
U-L.\ Pen, U.\ Seljak and N. Turok  
{\rm Phys.\ Rev.\ Lett} {\bf 79}, 2624 (1997). 

\bibitem{AveCal2} 
P.\ P.\ Avelino, R.\ R.\ Caldwell  
{\rm Phys.\ Rev.\  D} {\bf 53}, 5339 (1996). 
 
\bibitem{AveShe5} 
P.\ P.\ Avelino, E.\ P.\ S.\ Shellard, J.\ H.\ P.\ Wu, B.\ Allen, {\rm in 
preparation}.  

\bibitem{Per}  
S.\ Perlmutter, et al., 1998, {\rm Nature}, {\bf 391}, 51.

\bibitem{Lin} 
C.\ H. Lineweaver, {\rm Astro-ph/9805326}. 

\bibitem{AveCal1} 
P.\ P.\ Avelino, R.\ R.\ Caldwell, and C.\ J.\ A.\ P.\ Martins  
{\rm Phys.\ Rev.\  D} {\bf 56}, 4568 (1997). 
 
\bibitem{BRA} 
R.\ A.\ Battye, J.\ Robinson and A.\ Albrecht, {\rm Phys.\ Rev.\ Lett.}  
{\bf 80}, 4847 (1998). 
 
\bibitem{Mar}  
C.\ J.\ A.\ P.\ Martins, {\rm Phys.\ Rev. D} {\bf 55}, 5208 (1997).

\bibitem{VS90}  
S.\ Veeraraghavan and A.\ Stebbins, {\rm Ap. J.} {\bf 365}, 37 (1990). 

\bibitem{CarPre}
S.\ M.\ Carroll, W.\ H.\ Press and E.\ L.\ Turner, {\rm Annu.\ Rev.\
Astron.\ Astrophys.} {\bf 30}, 499 (1992).

\bibitem{Eis}
D.\ J.\ Eisenstein, {\rm Astro-ph/9709054}.

\bibitem{LahLil}
O.\ Lahav, P.\ B.\ Lilje, J.\ R.\  Primack and M.\ J.\ Rees, 
{\rm Mon.\ Not.\ R.\ Astron.\ Soc.} {\bf 251}, 128 (1991).

\bibitem{CheMag}  
C.\ Cheung, J.\ Magueijo, {\rm Phys.\ Rev.\ D} {\bf 56}, 1982 (1997).

\bibitem{AveCar}
P.\ P.\ Avelino and J.\ P.\ M.\ Carvalho, {\rm Mon.\ Not.\ R.\ Astron.\
Soc.} (1998).




\end{references}
\end{document}